\documentstyle[]{mn}

\title[The old stellar component of the  Galactic bar]
 {Detection of the old stellar component of the major Galactic bar }

\author[P.~L. Hammersley  et al.]{P.~L. Hammersley$^{1}$, F. Garz\'on$^{1,2}$, T. Mahoney$^{1}$, M. L\'opez-Corredoira$^{1}$,\\ \\
 {\LARGE M.~A. ~P. Torres $^{1}$}\\
$^1$ Instituto de Astrof\'\i sica de Canarias, E-38200  La Laguna,  Tenerife, Spain\\
$^2$ Departamento de Astrof\'\i sica, Universidad de La Laguna,  La Laguna,  Tenerife, Spain
}

\begin{document}

\maketitle

\begin{abstract}
We present near-IR colour--magnitude diagrams and star counts for a
number of regions along the Galactic plane. It is shown that along the
$l$=27$^\circ$ $b$=0$^\circ$ line of sight there is a feature at
5.7~$\pm~0.7$~kpc with a  density of stars  at least a factor two and
probably more than a factor five times that of the disc at the same
position. This feature forms a distinct clump on an $H$ vs. $J-H$
diagram and is seen at all longitudes  from the bulge to about
$l$=28$^\circ$, but at no longitude greater than this. The
distance to the feature at $l$=20$^\circ$ is about 0.5~kpc further than
at $l$=27$^\circ$ and by $l$=10$^\circ$ it has merged with, or has
become, the bulge.  Given that at $l$=27$^\circ$ and $l$=21$^\circ$
there is also a clustering of very young stars, the only component that
can reasonably explain what is seen is a  bar with half length of
around 4~kpc and a position angle of about 43$^\circ\pm 7^\circ$.

\end{abstract}


\section{Introduction}
The position of the Sun in the Galactic disc makes it difficult to
study the large-scale  structures in the inner Galaxy. The interstellar
extinction means that visible wavelengths cannot
penetrate more than a few kpc into the plane, so  until large-area IR
surface brightness maps and point source surveys became available little
could be said on the distribution of stars in the inner regions. Even
in the infrared  there remains the problem that a single line of sight will
contain sources from many Galactic components, which makes the
results ambiguous. Even the Galaxy's Hubble type, of importance in
 understanding how the Milky Way relates to other galaxies, is
still uncertain.

 Recently it has become accepted that the Galaxy is a barred spiral,
however there is still some considerable argument over what the bar
actually is.  There is general agreement  that the near end of the
Galactic bar is in the first quadrant, but angles between the bar and
the line of sight between the Sun and the Galactic Centre (GC) vary between about
10$^\circ$ and 75$^\circ$. The most popular angle based on studies of
the distribution of stars is between 10$^\circ$  and 30$^\circ$ (Dwek 1995;  Freudenreich 1998; L\'opez-Corredoira et al. 2000 ).  However, these studies 
are all based on off-plane bulge sources and so although the feature
is usually referred to as the bar, it could equally well be
described as a triaxial bulge. Binney et al. (1991) studied the flow of gas in the inner few degrees and also showed that the gas-flow lines followed a 
bar-like potential with a position angle of 16$^\circ$. 

 Other authors have studied other regions in
the plane and arrived at a larger angle for the bar. Sevenster et al. (1999) examined  
the kinematics of OH--IR stars and find a bar with an angle of 44$^\circ$. Peters (1975) on the basis of HI maps  and Nakai (1992) using large scale CO maps have both suggested a bar near 45$^\circ$ .
The most  
extreme bar angle yet suggested is 75$^\circ$ (Hammersley et al. 1994, hereafter Paper 1). This was based on the detection a large star formation region
between $l$=27$^\circ$ and 21$^\circ$ and then showing that a stick-like
bar with a half length of about 4~kpc  would readily explain the form of
the $COBE$ surface brightness maps in the Galactic plane.  Other
authors, however, have preferred to ascribe these features to a ring or blobs
of star formation in spiral arms or trailing from the triaxial bulge (e.g. Freudenreich 1998).

\begin{figure*}
 \centering
 \vspace{22.7cm}
\hspace{-22cm}\includegraphics{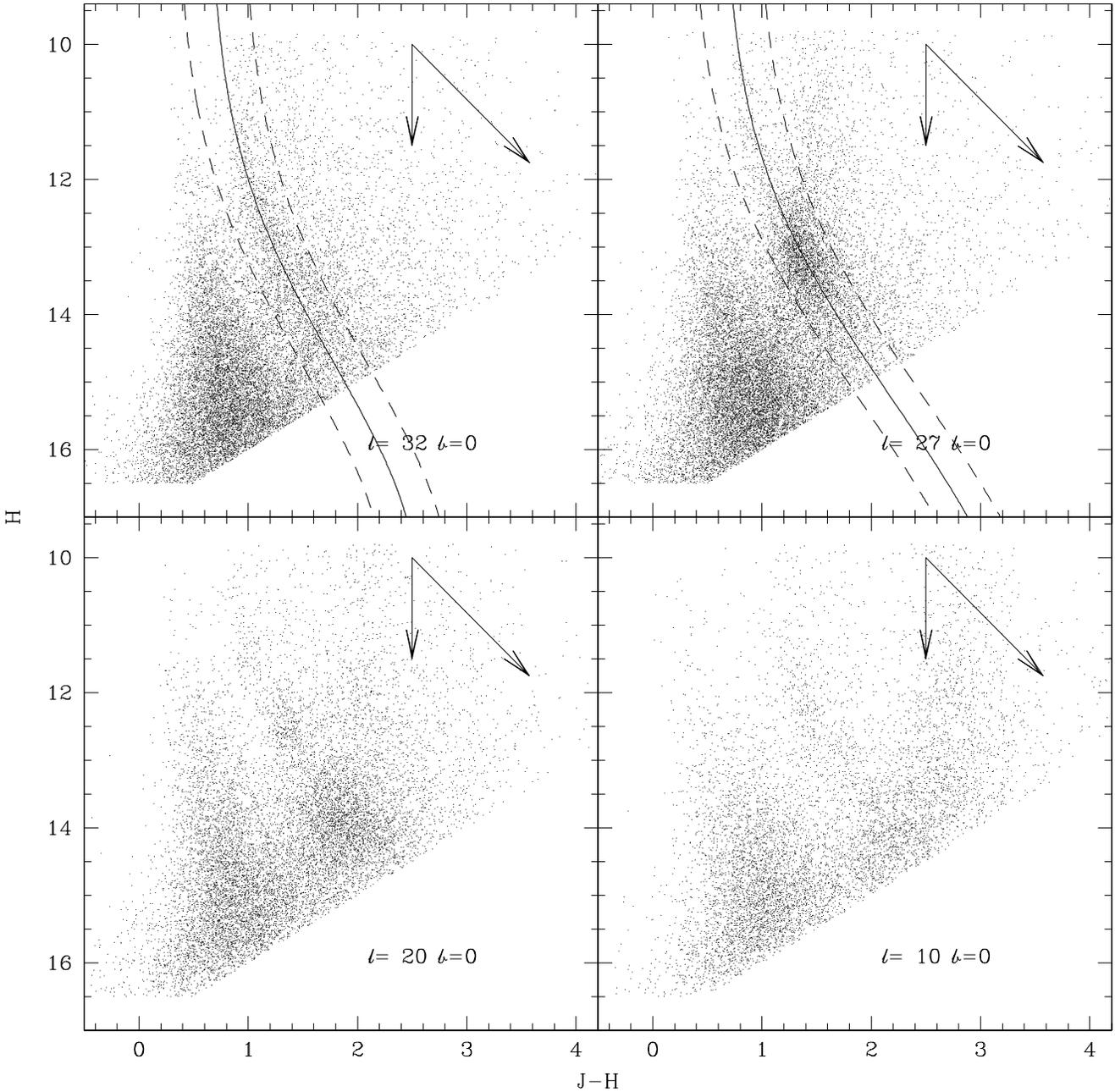}
 \vspace{-5.7cm}
 \caption{ The $J-H$ vs. $H$ diagrams for regions on the plane at
$l$=32$^\circ$, 27$^\circ$, 20$^\circ$ and 10$^\circ$.  The diagonal
arrow shows 10 magnitudes of visible extinction and the vertical shows
the effect of doubling the distance. The solid line going downwards
from left to right on the  $l$=32$^\circ$ and 27$^\circ$ plots show the
position of a K3III with increasing distance from the Sun.  The two almost vertical solid lines on the  $l$=27$^\circ$  plot  show giant branches at 
 4.5~kpc with an extinction of $A_V$=4.7mags and 6.6~kpc with $A_V$=8.7mags. }
 \label{f2*}
\end{figure*}

 If there really were a major bar, then as well as having
major SFRs at both extremes it could also have a  high concentration of older
stars along its whole length, if the bar is a long-lived feature. Such a clustering would form a strong
giant branch in a colour--magnitude diagram  for any line of sight
crossing the bar. The best way to search for an old population  deep
within the Galactic plane is to use near-IR  colour--magnitude diagrams.
Infrared wavelengths penetrate the interstellar dust far more easily than 
optical wavelengths. Furthermore, early  K giants,  which have by far the
highest space density for the giants, have a very restricted range of
absolute magnitudes. This means that they naturally form a dense clump
on an infrared HR diagram.  At the distance of interest these giants will
have a $K$ magnitude  of  12 to 14 making them simple to detect in
a few seconds on even a small telescope.

 In this paper we assume that the distance to the GC is 8~kpc, and that all positions quoted are on the Galactic plane unless otherwise stated.  We will also assume absolute magnitudes for K2-3III of $M_K$=$-$1.65 $M_H$=$-$1.5 and $M_J$=$-$0.89 as in Table 2 of  Wainscoat et al. (1992) and interstellar extinctions as given in Reike \& Lebofsky (1985) $A_K$=0.112$A_V$, $A_H$=0.175$A_V$ and $A_J$=0.282$A_V$.

\section{Observations}

During 1999 June observations of a series of  $20\times12$  arcmin fields
were observed along the Galactic plane between $l$=0$^\circ$ and $l$=37$^\circ$, using CAIN, the
facility IR camera on the 1.5~m TCS (Observatorio del Teide, Tenerife). The seeing was
typically 1$''$ and data were obtained only in photometric
conditions. Star counts were then produced to $J$=17, $H$=16.5 and $K_{\rm s}$=15.2.
Figure 1 show the $H$ vs. $J-H$ colour--magnitude diagrams for the fields on
the plane at $l$=32$^\circ$, 27$^\circ$, 20$^\circ$ and 10$^\circ$.  The
diagonal arrows shows the effect of 10 magnitudes of interstellar extinction in the visible.

\section{Analysis}

 The main sequence stars have relatively low intrinsic luminosities, so
only the closer ones  are detected and there is little interstellar
reddening. They form the triangular clump at $J-H$ = 0.5. To the
right of the main sequence  there is a diagonal curving stripe running from top left to bottom right.  This
is formed by the K giants in the disc with increasing distance from the
Sun. The middle solid line  running diagonally downwards from left to right on the $l$=32$^\circ$ and 27$^\circ$ plots,  shows
the position of a K3III star on this diagram. This assumes 
that the extinction at any point follows a double  exponential with a scale height of 50~pc and scale length of 3500~pc, as described in Wainscoat et al. (1992). 
A change in
absolute magnitude displaces the line vertically whereas a change in
the extinction changes the slope of the line in the diagram.  Therefore,
the fact that this line fits the stripe means that there cannot be
significant errors in the assumed absolute magnitudes and extinction.
The main sequence stars and the K-giant stripe can also be clearly seen
in the $l$=27$^\circ$, 20$^\circ$ and 10$^\circ$ plots. A cluster of
stars along the line of sight will form a giant branch extending
upwards and slightly to the right (e.g. the bulge at $l$=10$^\circ$ at
$J-H$=2.8). On the $l$=27$^\circ$ plot two giant branches have been plotted for
distances of 4.5kpc with an extinction of $A_V$=4.7mags and 6.6kpc with $A_V$=8.7mags.

In the middle of the $l$=27$^\circ$ stripe (m$_H$=13.3,$J-H$=1.3) there
is a large clump of sources with a giant branch extending almost
vertically to brighter magnitudes.  This clump can also be seen at
$l$=20$^\circ$ although somewhat more reddened ($J-H$=2) and this extra
extinction causes the giant branch to become less distinct.  Extinction
is patchy even on small angular scales and this leads to variations in
the extinction to the individual sources detected. Therefore, if the
average extinction rises the scatter in the magnitudes and colours will
also increase.  At $l$=10$^\circ$ the giant branch can be seen at
$J-H$=2.8, which is due to the bulge. The distance to the bulge plus
extinction is too great to see the  clump of K giants on the stripe.

  In order to gain an idea of the number of sources in the clump, the K
giants in the $l$=32$^\circ$ and 27$^\circ$ regions were isolated.  The
sources with a $J-H$ within 0.3 mag of the  predicted K3III line  were
extracted (the dashed lines in Fig. 1) and the $K_{\rm s}$ star counts
for only these source plotted in Figure 2.  $K_{\rm s}$ was used as it
is less effected by extinction than $H$ or $J$.  The sources used here
have  to be detected in all three bands and the limiting magnitude for
this plot is actually set by  $J$ and not $K_{\rm s}$,
which is why the limit is around 13.5 and not 15.2.  For magnitudes
brighter than about m$_{K_{\rm s}}$=11.5 there is little difference
between the two plots but then by m$_{K_{\rm s}}$=12.8 there are three times as many
sources at $l$=27$^\circ$.  The clump sources are extremely well
defined; the FWHM is under 1 magnitude and the spread that is seen can
be attributed principally to the luminosity function (LF) with a small part
to the differences in extinction and errors in the photometry. This
leaves very little spread due to the distance through the feature,
which implies that the feature is compact along the line of sight.  As
the LF is not precisely known, it is impossible to determine the distance through
the clump from these data alone, however it must be significantly less
than 2~kpc as  alone this would lead to a spread in magnitudes of 0.4
mag.

\begin{figure}
 \centering
 \vspace{11cm}
\hspace{-10cm}\includegraphics{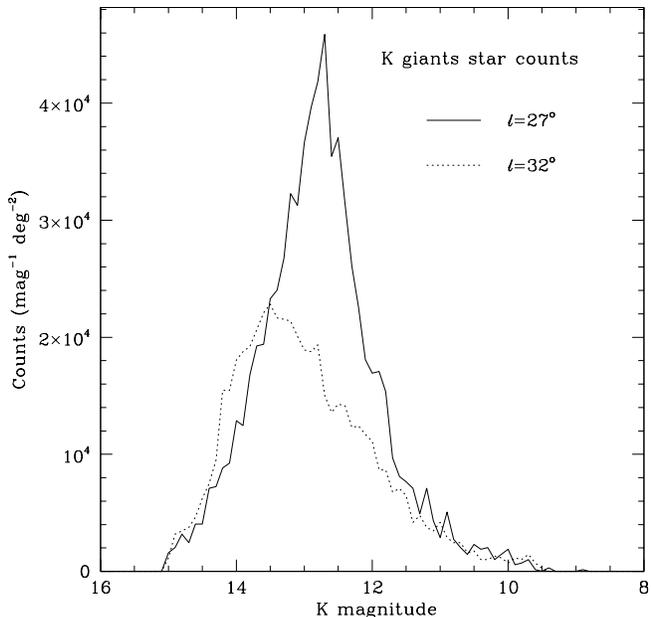}
 \vspace{-3cm}
 \caption{Differential $K_{\rm s}$ star counts for the K giants 
at $l$=32$^\circ$ and 27$^\circ$.}
\label{f3}
\end{figure}
\begin{figure}
 \centering
 \vspace{12cm}
\hspace{-10cm}\includegraphics{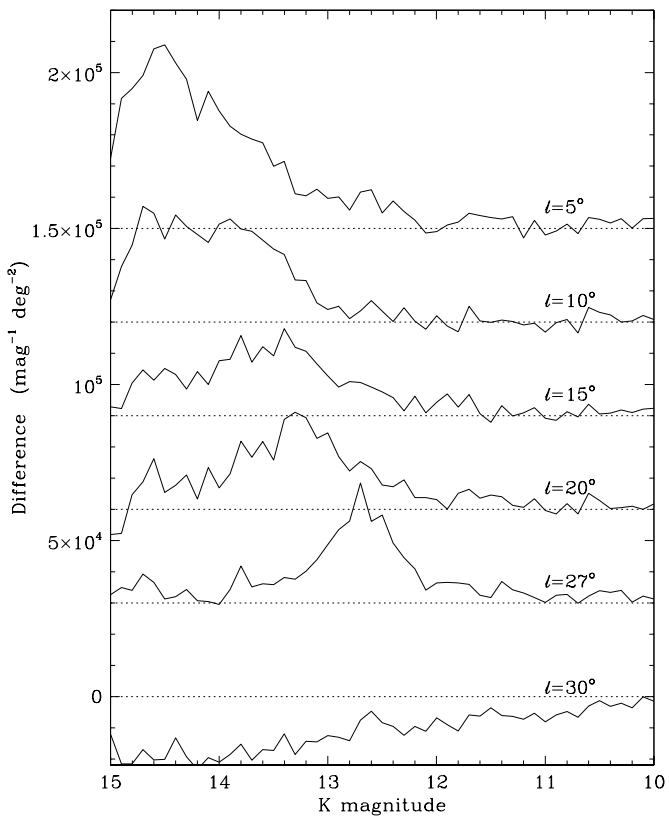}
 \vspace{-1cm}
 \caption{ Differential K$_{\rm s}$ star counts after removing the
dwarfs and subtracting the non-dwarf counts at $l$=32$^\circ$ for the
regions on the plane at $l$=30$^\circ$, 27$^\circ$, 20$^\circ$,
15$^\circ$, 10$^\circ$ and 5$^\circ$.  The curves are offset from each
other by 30,000 counts and the dotted horizontal lines mark the true
zero for each curve. }
 \label{f3}
\end{figure}

Figure 3 shows the $K_{\rm s}$ differential  star counts for six regions after
first removing the dwarfs and then subtracting the non-dwarf counts at
$l$=32$^\circ$ from the resulting counts in each region.  The  regions have been
offset by 30,000 counts to separate the plots and the zero for each
region  is shown by a dotted line.  The dwarfs were removed using
their position on the $H$ vs. $J-H$ diagram in order to increase the contrast
of any features in the inner Galaxy. There are no major clumps if the HR diagram at   $l$=32$^\circ$ and the  counts follow a smooth
curve with magnitude (Fig. 2). By subtracting the  non-dwarf $l$=32$^\circ$ counts, the steep rise in counts with increasing magnitude is reduced and the differences between the regions become far clearer. All the giants rather than just the K giants (as in Figure 2) were plotted as the $J$-band sensitivity is not sufficient to follow these stars to the distance of the bulge, also in many areas the K giant stripe becomes distorted by extinction. 

The line of sight at $l$=30$^\circ$ runs into the dust lane inside the
Scutum spiral arm whilst the $l$=32$^\circ$ region is tangential to the
stars in the arm. Therefore, the $l$=30$^\circ$ region has more extinction than
$l$=32$^\circ$ and hence the curve runs well below zero. In all of the
other regions the plots are more or less zero between m$_K{\rm
s}$=10 and 12, which  is surprising given the large difference in
longitudes, however it does indicate that the regions can be directly compared.
The peak at $l$=27$^\circ$, m$_{K_{\rm s}}$=12.8
shown in Figs. 2 and 3 can be clearly seen ($l$=27$^\circ$ was used
since at  $l$=25$^\circ$ there is a known dust lane which severely
distorts the counts). This peak is also  seen at $l$=20$^\circ$,
15$^\circ$, 10$^\circ$ and 5$^\circ$  but the magnitude of the peak
goes fainter as the longitude decreases.  At $l$=27$^\circ$, 20$^\circ$
and 15$^\circ$ the bulge should not provide any significant counts
(e.g. Freudenreich 1998; L\'opez-Corredoira et al.  2000), although by
$l$=10$^\circ$ and 5$^\circ$ the bulge is becoming important and this
is seen in the increased number of sources and the increased width of
the peak.  Therefore, the clump at $l$=27$^\circ$ cannot belong to
the bulge, but to a feature that runs into the bulge.

 The $J-H$ position of the clump (Fig. 1) is at 1.32 which indicates an
 extinction along the line of sight to the feature at $l$=27$^\circ$ of
 A$_V$=6 $\pm 1$ magnitudes. The peak of the K giants in Figure 3 is at
 m$_{K_{\rm s}}$=12.8 and assuming an absolute magnitude of $-1.65$
for the K giants and that $A_{K_{\rm s}}$=0.11$A_V$ gives the best
distance from the Sun to the feature at $l$=27$^\circ$ of  5.7$\pm$0.7~kpc.
 The error is dominated by uncertainty in the absolute magnitude and
extinction. The stated error was determined from the accuracy of the fit of the line to the peak of the K gaint stripe in Figure 1, which is 0.20 mags at the
distance of the cluster. At $l$=20$^\circ$ the peak at $K_{\rm s}$ is about 0.5
mag fainter   than at $l$=27$^\circ$ which, when extinction is taken
into account, makes it about 0.5~kpc further away (Figs. 1 and 3).

\begin{figure}
 \centering
 \vspace{11cm}
\hspace{-10cm}\includegraphics{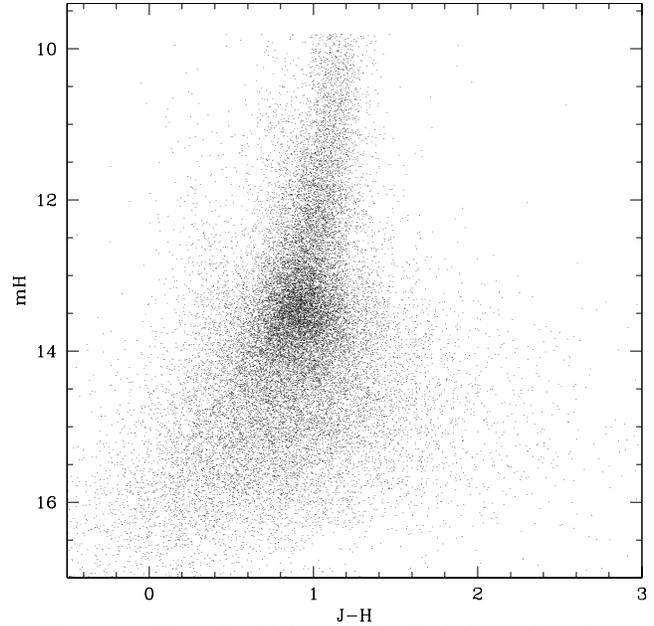}
 \vspace{-3cm}
 \caption{$H$ vs. $J-H$ diagram for the bulge at $l$=2$^\circ$ $b$=$-2^\circ$.}
\end{figure}

 As cross-check for the distance to the clump at $l$=27$^\circ$ Figure
4 shows the $H$ vs. $J-H$ diagram for the region at $l$=2$^\circ$
$b$=$-2^\circ$. The bulge is clearly seen forming a giant branch
extending upwards and slightly to the right near $J-H$ =1. However, as the
line of sight runs below the Galactic plane, the extinction is far less
than in Figure 1. The K giants again form a very strong clump at
m$_H$=13.5, $J-H$ =0.95. Therefore, assuming that these sources have a similar
absolute magnitude as the clump at $l$=27$^\circ$, this  implies that the
clump at  $l$=27$^\circ$  is at a distance of about 0.69 that of the bulge,
i.e. 5.5~kpc.  Conditions in the inner disc will be different from
those in the bulge, hence this calculation can only be approximate but
it is in agreement with the above  distance.

\begin{figure}
\centering
\vspace{11cm}
\hspace{-10cm}\includegraphics{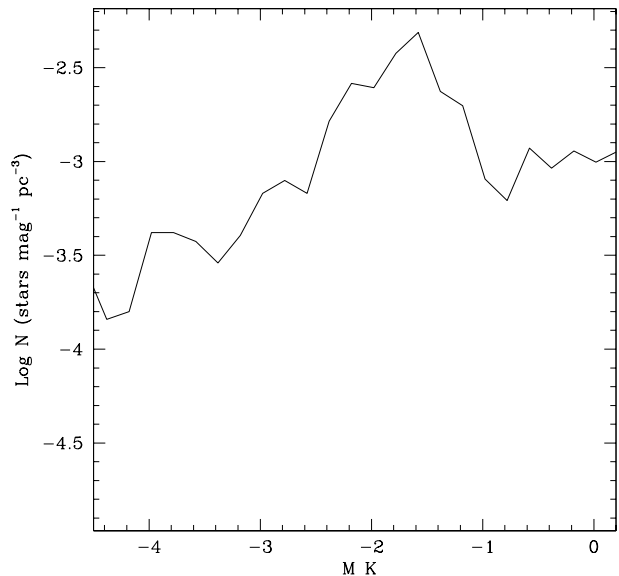}
\vspace{-3.2cm}
\caption{The K luminosity function of the clump at $l$=27$^\circ$, assuming a 
 distance through the clump of 500pc }
\label{f3}
\end{figure}

Figure 5 gives part of the luminosity function for the clump sources in
stars mag$^{-1}$pc$^{-3}$ assuming that the distance through the clump
is 500~pc, if the distance turns out to be larger then the density will
be proportionally less. It was obtained by isolating the sources
between the giant branches shown in the  $l$=27$^\circ$ plot in Figure
1. The sources were then  de-reddened assuming a distance of 5700~pc
and the standard extinction values,  the apparent magnitudes were then
converted into absolute magnitudes. The process was repeated for
$l$=32$^\circ$ and the result was then subtracted from that for
$l$=27$^\circ$ to remove the disc and leave the the clump sources. No
attempt has been made to correct for the distance through the clump or
error in the photometry. For comparison the disc  at the position of
the clump would have a density between 5 and 10 times lower  whereas
the bulge at this location would have a density over 100 times lower
(e.g. Wainscoat et al., 1992,  L\'opez-Corredoira et al. ,2000).  The
bulge reaches this sort of density at only a few hundred pc from the
GC.  Given that there is also a large number of very luminous young
sources at $l$=27$^\circ$ (Paper 1), then  this must be one of the most
luminous and densely populated  parts of the Galaxy  after the GC
itself.

\section{Discussion}
The $K_{\rm s}$ star counts clearly show a feature where the number of
old sources is similar between $l$=27$^\circ$ and $l$=5$^\circ$, but
whose distance from the Sun increases with decreasing longitude.  Paper
1 discusses the possible causes for the very high numbers of very
luminous stars also seen at $l$=27$^\circ$ and $l$=21$^\circ$ and shows that
these stars are related with the peaks in the COBE surface maps. Like the old star these young stars are seen at $l$=27$^\circ$  but not at $l$=32$^\circ$.
The distance to the clump of K giants at $l$=27$^\circ$ is about
5.7~kpc, which is very close to the value determined in Garz\'on et
al.  (1997) for the luminous stars  at $l$=27$^\circ$. This implies
that there is an old  population co-existing with, or very close to, a
young population.  Hence, the peaks in the $COBE$ surface brightness
maps near $l$=27$^\circ$, are important for understanding the
structures in the inner Galaxy and should not be dismissed as  patchy
star formation.

Paper 1 concludes that the most probable explanation  for the young stars seen at $l$=27$^\circ$ is a bar, but that a ring cannot be ruled out.
The above data  discounts the hypothesis that the old sources seen 
here  belong to a ring
\begin{itemize}
\item A ring, which is seen tangentially at $l$=27$^\circ$, should have far more sources at $l$=27$^\circ$ than at $l$=15$^\circ$, but this is not seen.
\item The ring would be significantly closer at $l$=15$^\circ$ than at 27$^\circ$, hence  the sources should appear brighter at $l$=15$^\circ$, not fainter  as seen here.
\end{itemize}
A long lived bar, however,  would naturally produce all of the features seen, in particular that the distance to the feature will increase with decreasing longitude if the near end of the bar is in the first quadrant.

  The distance to the clump at $l$=27$^\circ$ fixes the position angle
 at 43$^\circ\pm 7^\circ$ and a half-length of about 4~kpc.  This is
clearly at odds with Paper 1, which gives an angle of 75$^\circ$.  Such
a large bar angle would mean that the distance to the bar would be
almost the same between $l$=27$^\circ$ and $l$=5$^\circ$ so the peak in
Figure 4 would be more or less in the same position and not show the
systematic change that is seen.  The  75$^\circ$ position angle was
determined  principally on star formation region the far end of the bar
corresponding to a peak at $l$=$-$22$^\circ$. A 43$^\circ$ bar angle
would put the far end of the bar near $l$=$-$12$^\circ$ and indeed
there is a peak in $COBE$ 2.2 $\mu$m surface brightness maps at this
position (see Figs.  1 and 6 of Paper 1). It should be noted that the
peak in the $COBE$ 2.2 $\mu$m surface brightness maps at
$l$=$-$22$^\circ$ still needs to be explained and it is possible that
there is a ring-like structure with tangential points at
$l$=27$^\circ$ and  $l$=$-$22$^\circ$, however a discussion of this is
beyond the scope of this letter.

Many authors have examined the bulge off the plane and found that it is
triaxial (e.g.  Dwek 1995;  Freudenreich 1998; L\'opez-Corredoira et
al. 2000), however the position angle is small, typically about
15$^\circ$. This is about 30$^\circ$ different from the angle derived
here for the bar.  Although  feature seen at $l$=27$^\circ$  cannot be
the bulge as the latter does not give significant counts beyond about
$l$=12$^\circ$, the bar apparently does runs into the bulge near
$l$=12$^\circ$.  Therefore, although geometrically the two features are
distinct, currently it is not clear if the are dynamically linked  and
hence different aspects of the same phenomenon or where there are two
independent  bar-like features in the inner Galaxy, the triaxial bulge
which dominates the inner few hundred pc and a major bar which extends
to about 4~kpc.

\section{Conclusions}
There is a major old component in the $l$=27$^\circ$ line of sight at a
distance of about 5.7 kpc from the Sun. The component is seen at
$l$=20$^\circ$ and 15$^\circ$ and merges with the bulge inwards of
this.  The $l$=27$^\circ$ to 21$^\circ$ region is already known to have
a very high density of young stars and hence all of the expected
properties of a bar with a position angle of around 43$^\circ$ are
present. The distribution of  old stars is therefore very similar
to that suggested for  CO in Nakai (1992). The position angle  is,
however, significantly different from that of   the triaxial bulge
leading to the possibility that the Milky Way is a double-barred spiral
galaxy.

\section{Acknowledgments}

The TCS is operated on the island of Tenerife by the Instituto de
Astrof\'{\i}sica de Canarias at the Spanish Observatorio del Teide of
the Instituto de Astrof\'{\i}sica de Canarias. We would like to thank 
the referee for some valuable suggestions.

\end{document}